\documentclass[ reprint,
 amsmath,amssymb,
 aps,
 prl
]{revtex4-2}
\usepackage[utf8]{inputenc}
\usepackage{amsmath}
\usepackage{xcolor}
\usepackage{graphicx}
\usepackage[colorlinks=true,allcolors=blue]{hyperref}
\usepackage{gensymb}
\usepackage{physics}
\usepackage{siunitx}
\usepackage{float}

\usepackage[normalem]{ulem}

\begin{document}
\title{Directly imaging spin polarons in a kinetically frustrated Hubbard system}
\author{Max L. Prichard$^{1*}$, Benjamin M. Spar$^{1*}$, Ivan Morera$^{2,3}$, Eugene Demler$^{4}$, Zoe Z. Yan$^{1,5}$ and Waseem S. Bakr$^{1,\dag}$}

\affiliation{$^1$ Department of Physics, Princeton University, Princeton, New Jersey 08544, USA\\
$^2$Departament de Física Quàntica i Astrofísica, Facultat de Física, Universitat de Barcelona, E-08028 Barcelona, Spain\\
$^3$ Institut de Ciències del Cosmos, Universitat de Barcelona, ICCUB, Martí i Franquès 1, E-08028 Barcelona, Spain\\
$^4$ Institute for Theoretical Physics, ETH Z\"urich, CH-8093 Z\"urich, Switzerland\\
$^5$ James Franck Institute and Department of Physics, The University of Chicago, Chicago, IL 60637, USA
}
\date{\today}

\begin{abstract}
The emergence of quasiparticles in quantum many-body systems underlies the rich phenomenology in many strongly interacting materials. In the context of doped Mott insulators, magnetic polarons are quasiparticles that usually arise from an interplay between the kinetic energy of doped charge carriers and superexchange spin interactions~\cite{brinkman1970single,trugman1988interaction,kane1989motion,auerbach1991small,grusdt2018parton,koepsell2019Imaging, koepsell2020microscopic, ji2021dynamical}. However, in kinetically frustrated lattices, itinerant spin polarons -- bound states of a dopant and a spin-flip -- have been theoretically predicted even in the absence of superexchange coupling \hbox{\cite{Haerter2005, zhang2018, morera2021attraction,morera2023high,davydova2023itinerant,schlomer2023kinetic}}. Despite their important role in the theory of kinetic magnetism, a microscopic observation of these polarons is lacking. Here we directly image itinerant spin polarons in a triangular lattice Hubbard system realised with ultracold atoms, revealing enhanced antiferromagnetic correlations in the local environment of a hole dopant. In contrast, around a charge dopant, we find ferromagnetic correlations, a manifestation of the elusive Nagaoka effect~\cite{Nagaoka1966,white2001density}. We study the evolution of these correlations with interactions and doping, and use higher-order correlation functions to further elucidate the relative contributions of superexchange and kinetic mechanisms. The robustness of itinerant spin polarons at high temperature paves the way for exploring potential mechanisms for hole pairing and superconductivity in frustrated systems~\cite{zhang2018,morera2021attraction}. Furthermore, our work provides microscopic insights into related phenomena in triangular lattice moir\'{e} materials~\cite{tang2020simulation, foutty2023tunable,ciorciaro2023kinetic,tao2023}.
\end{abstract}

\maketitle

One of the key questions in quantum condensed matter physics is how doped Mott insulators give rise to exotic metallic and superconducting phases. Understanding this problem is crucial for explaining the emergence of the unusual physical properties of many families of strongly correlated electron systems, including the high-$T_c$ cuprates~\cite{keimer2015quantum}, organic charge transfer salts~\cite{powell2011quantum} and moir\'{e} materials~\cite{andrei2020graphene,mak2022semiconductor}. An important aspect of this problem is the interplay between spin order and the quantum dynamics of mobile dopants. So far, most studies have focused on Mott insulators on a square lattice where the motion of charge carriers disturbs spin correlations, resulting in an adversarial relationship between doping and spin order ~\cite{brinkman1970single,kane1989motion,auerbach1991small,grusdt2018parton,koepsell2019Imaging,  koepsell2020microscopic, ji2021dynamical}. This explains why many theoretical studies of doped high-$T_c$ cuprates are usually done from the perspective of Mott states in which spin order has been suppressed by fluctuations~\cite{georges1996dynamical,anderson2004physics,Lee_Nagaosa_Wen_2006}.

Recently, experiments on moir{\'e} materials have provided a strong motivation for understanding doped Mott insulators in triangular
lattices~\cite{mak2022semiconductor}. We explore this problem microscopically using a cold-atom triangular Fermi-Hubbard system~\cite{mongkolkiattichai2022quantum,xu2022doping}. One surprise of our experiments is that in contrast to square lattice systems, there is a symbiotic relation between mobile holes and antiferromagnetism. This manifests in the formation of antiferromagnetic (AFM) itinerant spin polarons in the hole-doped system, which we directly image by measuring spin correlations around mobile holes. In striking contrast, we find that particle doping favors the formation of ferromagnetic (FM) polarons similar to those discussed previously for the square lattice Fermi-Hubbard model~\cite{Nagaoka1966,white2001density}.

Some of the most important implications of our results are for systems in which the local interaction $U$ is much larger than the single electron tunnelling $t$, in which case the magnetic superexchange $J$ is strongly supressed. Indeed, this is the relevant regime for most moir\'{e} systems (neglecting  nearest-neighbor interactions). Intuition based on earlier studies would suggest that at temperatures higher than the superexchange scale, the regime we explore here, one can not expect coherent propagation of quasiparticles \cite{brinkman1970single}. Our results demonstrate that this does not have to be the case in triangular lattices. Formation of polarons around mobile dopants facilitates their propagation and makes their dynamics more coherent. This robustness of the quasiparticle can also be understood as the result of effective magnetic interactions with energy scale $t$ induced by the motion of dopants in the frustrated system~\cite{Haerter2005}.  

\section{Itinerant Spin Polaron}

\begin{figure*}[t]
\includegraphics[width = \textwidth]{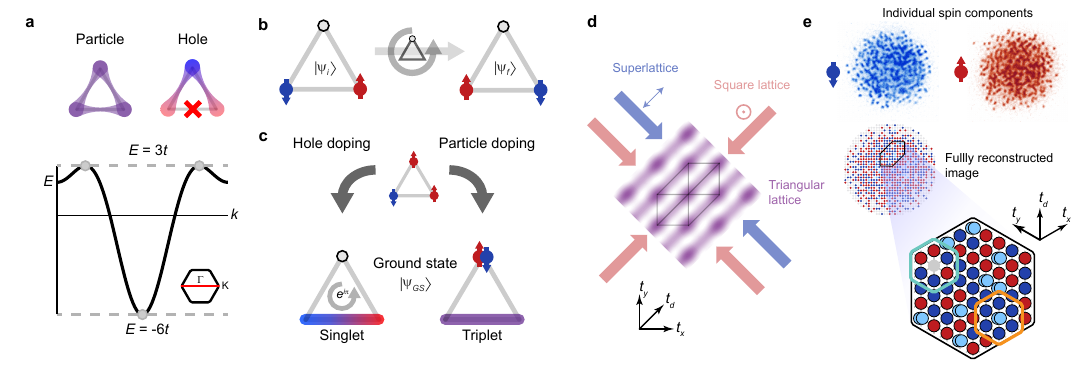}
\caption{\label{fig:1}
\textbf{Itinerant spin polaron.} \textbf{a,} A single particle in a triangular lattice with $t > 0$ minimizes its energy by occupying symmetric orbitals on each bond. Its band structure $E(k)$ exhibits a minimum energy of $E=-6t$. In a spin polarized background, a single hole has a negative effective tunneling and tries to occupy asymmetric orbitals on each bond, but is unable to do so (kinetic frustration). Its minimum energy is $E=-3t$. \textbf{b,} Once spins of the background Mott insulator are considered, the motion of a hole in a closed loop on a plaquette exchanges the spins. If the neighboring spins are in the singlet ($S = 0$) sector, the final state $\ket{\psi_f}$ picks up a spin Berry phase, i.e. $\ket{\psi_f} = e^{i \pi} \ket{\psi_i}$. This phase is absent in the triplet ($S=1$) sector. \textbf{c,} The relative sign flip for hole (particle) dopants means that a spin singlet (triplet) configuration is favored, manifesting as an AFM (FM) polaron. \textbf{d,} An optical lattice with triangular connectivity is formed by superimposing non-interfering square and 1D lattice potentials. The nearest-neighbor tunneling matrix elements are indicated as $t_x$, $t_y$ and $t_d$. \textbf{e,} A single fluoresence image gives the spatial distribution of both spin states. The reconstructed image contains hole (gray) and particle (light blue) dopants in a Mott insulator surrounded by either ferromagnetically or antiferromagnetically correlated spins. Spatial distances in the highlighted region of the lattice have been transformed to reflect the connectivity of the lattice.}
\end{figure*}

At the heart of the mechanism responsible for the formation of polarons in our experiment is the phenomenon of kinetic frustration, which has received much recent theoretical attention \hbox{\cite{zhang2018, morera2021attraction,chen2022proposal,morera2023high,davydova2023itinerant,Samajdar_Bhatt_2023,schlomer2023kinetic,lee2023triangular,van2022holes}}. This describes the reduction of the mobility of dopants due to the destructive interference of different propagation paths in certain lattice geometries, including the triangular one. To release this frustration and lower the kinetic energy of the dopants, the system develops magnetic correlations. The resulting magnetism, known as kinetic magnetism, is closely related to that studied by Nagaoka is his seminal work~\cite{Nagaoka1966}, but is more robust in that it is predicted to survive for finite interactions and doping. 

Indeed, signatures of kinetic magnetism above half filling have been observed very recently in doped van der Waals heterostructures through measurements of the spin susceptibility~\cite{tang2020simulation,ciorciaro2023kinetic}, while separate measurements of magnetisation plateaus attributed to kinetic effects below half filling have also been measured in these materials~\cite{tao2023}. Our experimental results provide a microscopic picture underlying these observations. More broadly, our results motivate studying the properties of doped Mott insulating states in triangular lattices, including superconductivity, from the perspective of self-organisation of itinerant spin polarons~\cite{alexandrov1996polarons,Lee_Nagaosa_Wen_2006,bohrdt2021exploration,morera2021attraction}.

\begin{figure*}[]
\includegraphics[width = \textwidth]{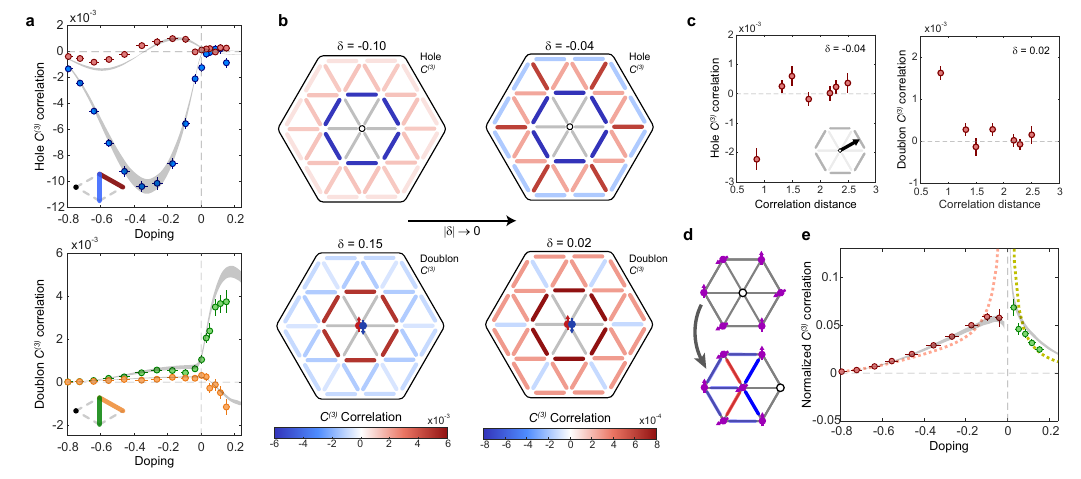}
\caption{\label{fig:2}
\textbf{Polaron internal structure vs doping.} \textbf{a,} Three point correlations $C^{(3)} ((1,0),(1/2, \sqrt{3}/2))$ and $C^{(3)} ((1,0),(3/2, \sqrt{3}/2))$ for holes (blue and red resp.) and doublons (green and orange resp.) versus doping $\delta$. Theory curves (gray bands) are from DQMC with $U/t = 11.8(4), T/t = 0.94(4)$.  
\textbf{b,}  Spatial map of correlations $C^{(3)}_h$ at $\delta = -0.10(2)$ and $\delta = -0.04(2)$ (top) and $C^{(3)}_d$ at $\delta = 0.15(2)$ and $\delta = 0.02(2)$ (bottom). The second column highlights the spatial structure of the polarons as close as we can get experimentally to the ideal single dopant limit.
\textbf{c,} Correlations $C^{(3)}_h$ and $C^{(3)}_d$ vs distance at dopings of $\delta = -0.04(2)$ and $\delta = 0.02(2)$, respectively. Corelation distance is defined as the distance from the dopant to the bond midpoint, shown as the inset black arrow. Bonds are unity length, implying the closest correlation is at a distance $\sqrt{3}/2$.
\textbf{d,} Dopant propagation coherently re-orders the surrounding 120$^\circ$ order, resulting in an alternating pattern of correlations, as imaged in the experiment.
\textbf{e,} Nearest-neighbor three-point correlators normalized by the dopant density. 
DQMC theory is shown for the interacting systems (gray bands) and in the non-interacting 
limit (dotted red and green lines). Error bars represent 1 standard error of the mean (s.e.m.).}
\end{figure*}

Our system consists of a two-dimensional degenerate gas of $^6$Li that is an equal mixture of two spin species corresponding to the first ($\ket{\uparrow}$) and third ($\ket{\downarrow}$) lowest hyperfine states of the atom. The gas is loaded adiabatically into an optical lattice realising the triangular lattice Hubbard model,
\begin{equation}\label{eq:1}
    \displaystyle {\hat {H}}=-\sum _{\expval{i,j},\sigma } t \left({\hat {c}}_{i\sigma }^{\dagger }{\hat {c}}_{j\sigma }+{\hat {c}}_{j\sigma }^{\dagger }{\hat {c}}_{i\sigma }\right)+U\sum _{i}{\hat {n}}_{i\uparrow }{\hat {n}}_{i\downarrow },
\end{equation}
where $ \hat{c}^{\dagger}_{i\sigma} $ ($\hat{c}_{i\sigma}$) creates (destroys) a fermion of spin at lattice site $i$, the number operator $\hat{n}_{i\sigma} = \hat{c}^{\dagger}_{i\sigma}\hat{c}_{i\sigma}$ measures site occupation and $\expval{i,j}$ denotes nearest-neighbor sites. In the model, particles hop with $t>0$. With this sign of the tunneling, a particle in an empty lattice can lower its energy by delocalizing on each lattice bond in a symmetric spatial orbital (Fig.~\ref{fig:1}a). The corresponding band structure is particle-hole asymmetric, and the particle attains its minimal energy of $-6t$ at zero quasi-momentum. Kinetic frustration can be understood by considering the opposite scenario of a single hole moving in a spin-polarized background. In this case, the Hamiltonian is better expressed in terms of hole operators with $\hat{h}_{i}^\dagger=\hat{c}_{i}$,

\begin{equation}\label{eq:2}
    \displaystyle {\hat {H}}=\sum _{\expval{i,j} } t \left({\hat {h}}_{i }^{\dagger }{\hat {h}}_{j }+{\hat {h}}_{j }^{\dagger }{\hat {h}}_{i }\right)
\end{equation}

Crucially, the change in the sign of the tunneling resulting from the anticommutation of fermionic hole operators in the Hamiltonian favors antisymmetric spatial orbitals for the hole on each bond.  Indeed, this is manifested in the band structure of the hole, which is mirrored about zero energy relative to the particle. The hole kinetic energy is thus minimized at a value of $-3t$, larger than in the unfrustrated system.

A simplified picture explaining the emergence of the itinerant spin polaron in the doped interacting system can be obtained by considering a triangular plaquette with two fermions. In the limit of strong interactions, double occupancies are energetically forbidden and the motion of the hole on a closed loop on the plaquette will exchange the two spins. In the spin singlet sector, this produces a spin Berry phase of $\pi$, whereas the phase is zero in the spin-symmetric triplet sector (Fig.~\ref{fig:1}b)~\cite{sposetti2014classical}. The phase acquired by the hole in the singlet sector returns the tunneling to a positive value, thereby releasing the kinetic frustration and allowing the hole to reach a lower ground state energy. 

The resulting object, a singlet bond bound to a hole with a binding energy of order $t$, is predicted to persist in the many-body setting for light hole doping (Fig.~\ref{fig:1}c). This corresponds to a polaron with antiferromagnetic spin correlations in the vicinity of a hole. The situation is reversed for particle doping, favoring ferromagnetic correlations in the vicinity of a doublon. We directly detect the itinerant spin polaron in our system using a connected three-point correlation function, which probes the spin correlations in the environment of a hole or doublon. Such correlators have been previously used to identify magnetic polarons in the square lattice, although in that case, the mechanism that leads to the formation of the polaron is different and the binding energy is on the superexchange scale~\cite{koepsell2019Imaging,ji2021dynamical,hartke2023direct}.

\begin{figure*}[]
\includegraphics[width = \textwidth]{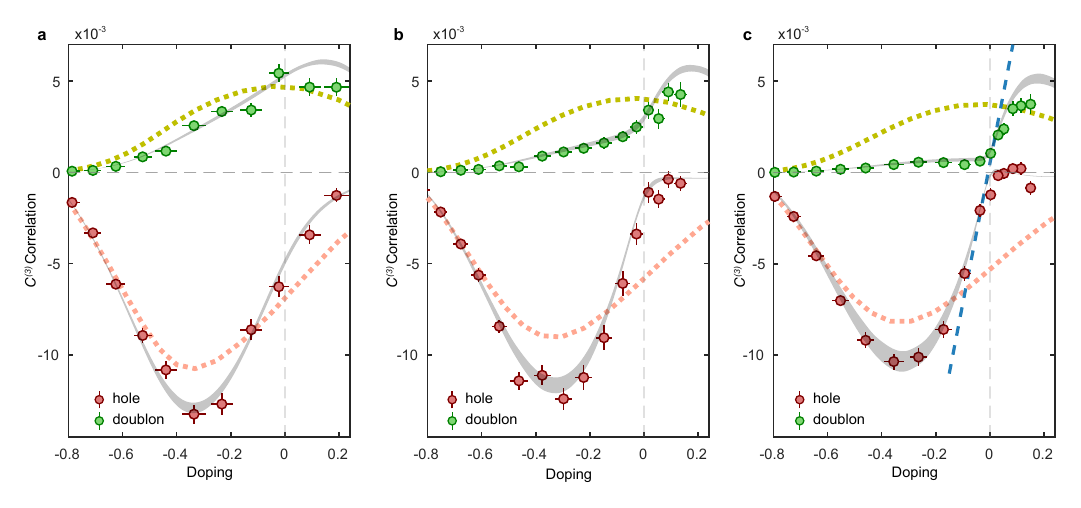}
\caption{\label{fig:3} \textbf{Evolution of three-point correlations with doping and interactions.} $C^{(3)}_h$  and $C^{(3)}_d$  connected correlations vs. doping in the metallic and Mott insulating regimes for $\mathbf{d_1}=(1,0)$, $\mathbf{d_2}=(1/2,\sqrt{3}/2)$. Data shown for \textbf{a,} $U/t= 4.4(1), T/t = 0.68(2)$, \textbf{b,} $U/t = 8.0(2), T/t = 0.84(3)$, and \textbf{c,} $U/t = 11.8(4), T/t = 0.94(4)$. For all interactions, we observe the largest negative $C^{(3)}_h$ correlator at a doping of around -0.3. For increasing $U/t$, the $C^{(3)}_h$  and $C^{(3)}_d$ correlators become more linear near zero doping, indicating a region where there are weakly interacting polarons. The blue dashed line in \textbf{c}, which is fit to the DQMC in the doping range $-0.09<\delta<0.06$, illustrates this region. DQMC theory is shown for the interacting systems (gray bands) and in the non-interacting limit (dotted red and green lines). Error bars represent 1 s.e.m.}
\end{figure*}

To realise a lattice with triangular connectivity ~\cite{yang2021site,mongkolkiattichai2022quantum,xu2022doping,Yamamoto_2020,Struck2011, wei2023observation, trisnadi2022design}, we superimpose two non-interfering lattices, a strong one-dimensional optical lattice with spacing $a = 532$ nm and depth $V_{532} = 6.7(2) \, E_{R,532}$ and a weak square optical lattice with larger spacing $a = 752$ nm and depth $V_{752} = 2.9(1) \, E_{R,752}$, where $E_{R,a} \equiv h^2/8ma^2$ with $m$ the mass of the atom (Fig.~\ref{fig:1}d)~\cite{wei2023observation}. 
The frequency detuning between the two lattices is used to tune their relative alignment to obtain a triangular geometry. Their relative depths are chosen to produce an isotropic triangular lattice by equalising the tunneling strength along the original square lattices axes and one diagonal. The gas is prepared at a magnetic field near a Feshbach resonance at 690 G allowing us to freely tune the scattering length. In this way we tune the coupling strength $U/t$ to explore the evolution of the correlations from the metallic to the Mott insulating regime. 

We use a quantum gas microscope to measure site-resolved correlations associated with the polaron in the many-body system~\cite{gross2021quantum}. We further implement a bilayer imaging technique~\cite{preiss2015quantum,hartke2020doublon,koepsell2020robust,Yan2022}, wherein a magnetic field gradient is first used to separate the two spin states into different layers prior to imaging them simultaneously (Fig \ref{fig:1}e). From the reconstructed images, we can calculate arbitrary $n$-point correlation functions involving both spin and density operators averaged over experimental cycles. In the strongly interacting regime, the atoms order in a Mott insulator and exhibit short-range 120$\degree$ spiral AFM correlations that have been observed in previous experiments~\cite{mongkolkiattichai2022quantum,xu2022doping}. We use the two-point spin correlations for thermometry by comparison to Determinant Quantum Monte Carlo (DQMC) calculations~\cite{varney2009quantum} (see Methods). The typical peak density of the clouds in the lattice is $n=n_{\uparrow}+n_{\downarrow}=1.2$, allowing us to study a range of dopings $\delta = n -1$ on either side of half-filling of the Hubbard system in each experimental snapshot due to the harmonic confinement of the lattice beams. 

\section*{Spin Polaron Correlations}
To detect the polaron, we evaluate connected three-point charge-spin-spin correlation functions. For a hole dopant, the relevant correlation function, $C^{(3)}_{h}(\mathbf{d_1},\mathbf{d_2})$ is computed as:
\begin{equation}\label{eq:3}
    C^{(3)}_{h}(\mathbf{d_1},\mathbf{d_2}) \equiv \langle \hat{n}^h_\mathbf{r_0} \hat{S}^z_\mathbf{r_0+d_1} \hat{S}^z_\mathbf{r_0+d_2} \rangle - \langle \hat{n}^h_\mathbf{r_0} \rangle \langle \hat{S}^z_\mathbf{r_0+d_1} \hat{S}^z_\mathbf{r_0+d_2} \rangle,
\end{equation}
where we have assumed a spin-balanced system $\langle \hat{S}^z_\mathbf{r}\rangle = 0$. Here $\mathbf{d_1}$ and $\mathbf{d_2}$ are displacement vectors relative to a site at position $\mathbf{r_0}$, $\hat{S}^z_\mathbf{r}=\hat{n}_{\mathbf{r}\uparrow}-\hat{n}_{\mathbf{r}\downarrow}$ is the projection of the spin on site $\mathbf{r}$ along the quantization axis and $\hat{n}^h_\mathbf{r}$ is the hole number operator $(1-\hat{n}_{\mathbf{r}\uparrow})(1-\hat{n}_{\mathbf{r}\downarrow})$. We average the correlation function over sites $\mathbf{r_0}$ with similar doping. The doublon correlation function $C^{(3)}_{d}$ is constructed in an analogous way by replacing the hole number operator $\hat{n}^h_\mathbf{r}$ with the doublon number operator $\hat{n}^d_\mathbf{r}=\hat{n}_{\mathbf{r}\uparrow}\hat{n}_{\mathbf{r}\downarrow}$. For the range of dopings we consider, the two-point spin correlator is always negative due to the dominant superexchange aniferromagnetism. The second term of the connected three-point correlators defined in equation (\ref{eq:3}) removes any uncorrelated charge-spin-spin signal associated with this background AFM signal.

We start by studying the doping dependence of $C^{(3)}$ above and below half filling, shown in Fig.~{\ref{fig:2}}a. Correlations $C^{(3)}_h$ ($C^{(3)}_d$) in the vicinity of holes (doublons) are shown for a strongly interacting sample at $U/t=11.8(4)$. For $\delta < 0$, $C^{(3)}_h$ exhibits negative correlations for $\mathbf{d_1}=(1,0)$, $\mathbf{d_2}=(1/2,\sqrt{3}/2)$,  indicating an enhancement of AFM order in the immediate vicinity of a hole. In contrast, for $\delta > 0$ the doublon correlation function $C^{(3)}_d$ is positive for the same bond, revealing a preference for ferromagnetic order around doublon dopants. We then examine correlations out to further distances to explore the structure of the polaron at a few different dopings in Fig.~{\ref{fig:2}}b. In particular, we emphasise the structure of the correlations in the limit of vanishing doping in the right column of this panel. This doping regime is the closest to capturing the behaviour associated with the idealized case of a single dopant \mbox{\cite{Haerter2005,Nagaoka1966}} where polaron-polaron interactions are absent. Around a hole (Fig.~{\ref{fig:2}}b,c), we find that while at the shortest distance $C^{(3)}_h$ is negative, the next furthest ring (distances 1.32, 1.5) shows positive correlations. The structure of the AFM hole polaron can be understood using a picture of a mobile hole dopant that coherently modifies the surrounding \mbox{120$^\circ$} N\'{e}el order to facilitate lowering the kinetic energy, as illustrated in Fig.~{\ref{fig:2}}d. Strikingly, this is a different structure than on the doublon side, where predominantly positive correlations exist up a distance $d\sim1.8$ away from the doublon. This indicates an energetic preference towards a locally ferromagnetic environment, as for a particle dopant the motion is unfrustrated in a background of polarized spins. We interpret these short-range polaronic correlations we observe as the precursors to Haerter-Shastry AFM and Nagaoka FM expected at lower temperatures on the hole- and particle-doped sides respectively.

We also note from Fig.~\ref{fig:2}a the approximate linear dependence of the correlators with relevant dopant density $\delta$ for $|\delta| \lesssim 0.1$. This indicates that in this regime, the description of the system in terms of weakly interacting polarons is valid. Polaron interactions become important for larger dopant densities. These observations motivate introducing a normalized version of the correlators by dividing out the relevant dopant density $\delta$, where $C^{(3)}_{\mathrm{norm}} \equiv C^{(3)}/\delta$ (Fig.~\ref{fig:2}e). While the non-normalized correlations close to zero doping appear reduced in magnitude, the normalized spin correlation emphasize the fact that the spin correlations per dopant are in fact strongest close to half-filling.

The itinerant spin polaron picture we have presented so far is in the regime of strong interactions, but it is also interesting to explore how the three-point correlations evolve with $U/t$. Fig.~\ref{fig:3} shows these correlations in the metallic ($U/t = 4.4$), Mott insulating ($U/t =11.8$) and intermediate regimes ($U/t = 8.0$) in the temperature range $T/t \sim 0.7-0.9$. Surprisingly, many of the qualitative features of the correlations are similar, including the minimum in the antiferromagnetic correlations around a hole at $\delta \sim -0.3$. This can again be understood from a single plaquette in the alternative limit of vanishing interactions, which predicts correlations of the same sign as the itinerant spin polaron \cite{Merino_Powell_McKenzie_2006} (see Methods). For all interactions, the measured correlations shows reasonable agreement with DQMC calculations with a systematic deviation in $C^{(3)}_d$ and $C^{(3)}_h$ for larger fillings, possibly due to an increase in reconstruction errors (see Methods). The correlations differ significantly from those expected for the non-interacting gas, especially for the two stronger interactions. As $U/t$ increases, the onset of the correlation moves closer to half-filling as contributions from virtual doublon-hole fluctuations are increasingly suppressed. The characteristic linear growth of the correlations, expected in the polaronic regime and observed for $U/t = 11.8$, is however only present for the strongest interactions. Additional evidence for an experimental observation of the itinerant spin polaron at the largest interaction strength comes by combining the observed three-point correlators with a measurement of the singles fraction $n_s = n-n^d$ at half-filling, which ensures that the system is in the strongly interacting regime. For $U/t = 4.4, 8.0$ and $11.8$, this is $0.71(1), 0.85(1)$ and $0.93(1)$ respectively.

\begin{figure}[]
\includegraphics[width=\columnwidth]{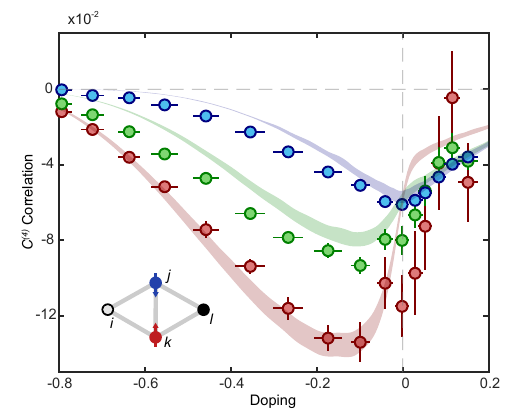}
\caption{\label{fig:4} \textbf{Comparing antiferromagnetic correlations
due to kinetic and superexchange mechanisms.} Conditional four-point correlators $C^{(4)}$ show the spin-spin correlator on the $j-k$ bond when surrounded by two holes (red), one hole and one singly occupied site (green), and two singly occupied sites (blue). Below zero doping, bonds have increased antiferromagnetic spin correlations with increasing neighbors that are holes. DQMC theory shown as colored bands. Error bars represent 1 s.e.m.}
\end{figure}

\section*{Kinetic vs. Superexchange Correlations}

Since superexchange-induced AFM correlations are present in the system for any finite interaction strength,  it is illuminating to quantify their strength in comparison to kinetic magnetism around dopants. This can be done using four-point correlation functions. For any two nearest-neighbour sites $j$ and $k$, there are two additional sites $i$ and $l$ (which we call conditioning sites) coupled to both of them (Fig.~\ref{fig:4} inset). As a dopant on either conditioning site may affect the correlation strength between $j$ and $k$, a four-point correlation function is required to determine the influence of the background on the $j-k$ bond. We define a conditional four-point correlator as the spin correlator on the shared bond, conditioned on the occupancy observables $\hat{n}^a, \hat{n}^b$ on sites $i$ and $l$,

\begin{gather}
    C^{(4)}_{ab} \equiv \frac{\langle\hat{n}^a_i \hat{S}^z_j \hat{S}^z_k \hat{n}^b_l \rangle}{\langle\hat{n}^a_i \hat{n}^b_l \rangle}  ,
\end{gather}
where the labels $a, b \in \{h,s\}$. Fig.~\ref{fig:4} shows the four-point correlator at $U/t=11.8$ for three different occupancies of the conditioning sites. Below half-filling, we find that $C^{(4)}_{hh}<C^{(4)}_{hs}<C^{(4)}_{ss}$, directly indicating the enhancement of AFM correlations on a given bond by the presence of holes on the conditioning sites. Kinetic magnetism, therefore, strengthens the existing AFM correlations below zero doping that arise due to superexchange, which is the only mechanism at play on the level of the four-site plaquette when the conditioning sites are singly-occupied. Above zero doping, where holes are due to virtual fluctuations, there is no enhancement of anti-ferromagnetic correlations from kinetic magnetism and all three correlators are similar in value. 

\section*{Outlook}

In this work, we have directly imaged itinerant spin polarons in a triangular Hubbard system by measuring three- and four-point correlation functions. We have characterised their evolution with doping and interactions, and compared the strength of correlations induced by superexchange and kinetic effects. In future work, it would be interesting to study the polaron spectroscopically~\hbox{\cite{morera2023exploring}}, which would allow direct characterisation of its binding energy as well as its dispersion and effective mass. While the polarons we have focused on here are part of the physics of the doped triangular Hubbard model at high temperatures, pushing to lower temperatures would shed light on its rich ground state phase diagram. Theoretical work suggests this may include magnetically ordered phases as well as a quantum spin liquid with fractionalised excitations at intermediate interactions~\hbox{\cite{zhu2022doped,Szasz2020}}. Higher-order connected correlations may be useful in identifying more complex multi-particle bound states in the system~\hbox{\cite{morera2021attraction}}, which can lead to hole-pairing mechanisms and superconductivity at high temperatures~\hbox{\cite{schrieffer1988spin,zhang2018,venderley2019density,morera2021attraction,zhu2022doped,zampronio2023chiral}}.

\textbf{Acknowledgements:} We acknowledge Markus Greiner, David Huse, Rhine Samajdar, Lawrence Cheuk, Eun-Ah Kim, Daniel Khomskii, Annabelle Bohrdt, Fabian Grusdt, Henning Schl\"{o}mer and Gil Refael for helpful discussions. We also thank Siddarth Dandavate for early assistance in performing the DQMC simulations. The experimental work was supported by the NSF (grant no. 2110475), the David and Lucile Packard Foundation (grant no. 2016-65128) and the ONR (grant no. N00014-21-1-2646). M.L.P. acknowledges support from the  NSF Graduate Research Fellowship Program. E. D. acknowledges support from the ARO (grant no. W911NF-20-1-0163) and the SNSF (project 200021\_212899). I.M. acknowledges support from Grant No. PID2020-114626GB-I00 from the MICIN/AEI/10.13039/501100011033 and Secretaria d’Universitats i Recerca del Departament d’Empresa i Coneixement de la Generalitat de Catalunya, cofunded by the European Union Regional Development Fund within the ERDF Operational Program of Catalunya (Project No. QuantumCat, Ref. 001-P-001644).

\textbf{Author contributions:} E.D., I.M. and W.S.B. conceived the study and supervised the experiment. M.L.P., B.M.S. and Z.Z.Y. performed the experiments and analyzed the data. All authors contributed to writing the manuscript.

\textbf{Competing interests:} The authors declare no competing interests. 
\\

\noindent$^*$ These authors contributed equally to this work.\\
\noindent $^\dag$ Email: wbakr@princeton.edu
\bibliography{polaron}

\setcounter{figure}{0}
\setcounter{equation}{0}
\setcounter{section}{0}

\clearpage

\renewcommand{\thefigure}{S\arabic{figure}}
\renewcommand{\theHfigure}{S.\thefigure}
\renewcommand{\theequation}{S\arabic{equation}}
\renewcommand{\thetable}{S\arabic{table}}

\preprint{APS/123-QED}

\section*{Methods}

\subsection*{State preparation}
We use a degenerate mixture of hyperfine states $\ket{1}$ and $\ket{3}$, where $\ket{i}$ represents the $i^{\mathrm{th}}$ lowest energy level of the ground hyperfine manifold of atomic $^6$Li, to simulate the two-component Fermi-Hubbard model on a triangular lattice. State preparation of a degenerate Fermi gas prior to loading the science lattice largely proceeds as detailed in previous work \cite{brown2017spin}. After the final stage of evaporation we are left with a spin-balanced sample of approximately 400 atoms in each spin state. At this stage, the atoms are confined in a single layer of an accordion lattice, created with 532 nm light, with spacing $a_z = $ 3.6(3) $\mu$m and trap frequency $\omega_z = 2 \pi \times 16.4(2)$ kHz in the vertical ($z$) direction. The combined in-plane 2D lattices (see below) are then ramped to their final depths following a cubic spline trajectory in 100 ms. An additional 1070 nm optical dipole trap propagating along $z$ with waist $w_0 = 100$ $\mu$m is used to provide variable confinement in the  $x-y$ plane in the final science configuration. In particular, for strongly interacting samples, the reduced compressiblity necessitates greater confinement to achieve comparable densities. We worked at  magnetic fields ranging from 587(1) G to 612(1) G, where the scattering length varies between 330(15) and 945(30) Bohr radii respectively.

\subsection*{Triangular optical lattice}
\begin{figure}[]
\includegraphics[width = \columnwidth]{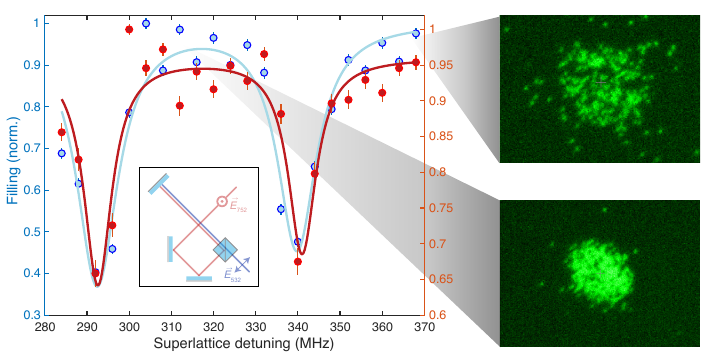}
\caption{\label{fig:532PBandSpec} \textbf{Superlattice Phase Calibration.} P-band spectroscopy used to calibrate the superlattice phase and stability. The dips at $292$ MHz and $340$ MHz superlattice detuning correspond to $\phi = 3 \pi/2$ and $\phi = \pi/2$. Fitted peaks (solid lines) are at [292.4(6) MHz , 339.8(8) MHz] and [292.6(8) MHz , 341.0(8) MHz] for the red and blue data respectively. Callout fluoresence images show the expansion of a Mott insulator after 1 second for superlattice phase $\phi = \pi$ (top) and $\phi = 0$ (bottom) at $V_{752} = 42.0(3) E_{R,752}$ and $V_{532} = 3.7(1) E_{R,532}$. Inset: Experimental setup used for realizing the optical superlattice.}
\end{figure}

The triangular lattice is formed as in \cite{wei2023observation} by combining two non-interfering lattices of different polarizations and detunings (Extended Data Fig. \ref{fig:532PBandSpec}, inset). Both lattices are created using light of wavelength 1064~nm. The first is a square lattice with a spacing of $752$ nm, created by retroreflecting a single vertically-polarized laser beam in a bowtie geometry. The depth of this lattice is calibrated using amplitude modulation spectroscopy. Both losses of laser power as the beam traverses the vacuum chamber and and non-orthogonal beam alignments can cause a significant tunneling imbalance along the axes of the square lattice. In our system, we specifically tune to $\theta = 90.7(1) \degree$ (measured using atomic fluorescence images) which approximately cancels the imbalance due to power losses. AFM correlations along the two axes of the square lattice show a systematic difference of $4(3)\%$, indicating a difference in the tunnelings of approximately $1(1)\%$.

The second lattice is a one-dimensional (1D) optical lattice with spacing $532$ nm and wavevector aligned with a diagonal of the $752$ nm square lattice. The light for this lattice is horizontally polarized and detuned by $\sim330$ MHz with respect to the square lattice, preventing any electric field interference. Both lattices share a common retroreflecting mirror, avoiding the need for active phase stabilization as in other schemes \cite{tarruell2012creating,xu2022doping}. 

The frequency detuning between the two lattices introduces a relative spatial phase between the two potentials at the atoms which is given by $\phi = 4 \pi L \Delta  / c$, where $\Delta$ is the relative detuning and $L$ is the distance from the atoms to the retroreflecting mirror. The triangular lattice configuration is obtained for the case of constructive interference, i.e. $\phi = 0$, which we calibrate using \emph{in-situ} measurements. The superlattice depth is set to a weak value ($V_{532} = 0.49(1)E_{R,532}$) relative to the dominant square lattice ($V_{752} = 40.2(3)E_{R,752}$) and then  modulated at the frequency of the square lattice $p$-band resonance. As this is an odd-parity transition, excitation should be maximized when $\phi = \pi/2$ or $3 \pi/2$ which induces a sloshing motion. The two corresponding prominent resonance peaks versus superlattice detuning at a constant modulation frequency are shown in Extended Data Fig. \ref{fig:532PBandSpec}. 

We perform two identical measurements separated by 1 week (red and blue data) to assess the long-term stability of the setup. The agreement of the resonance peaks between the two datasets is at or below the uncertainty ($\sim1$ MHz) of the spectroscopic measurement, indicating phase stability at or below $0.02 \pi$ radians. Explicit band structure calculations show that such a phase drift results in a negligible change of the tunneling values of less than $0.2$ Hz on top of a tunneling strength of $400$ Hz.

We note that the $\phi = 0$ (triangular) and $\phi = \pi$ (honeycomb) conditions are indistinguishable from spectroscopic measurements alone as they both produce an even parity drive. To distinguish these two phases, a dense Mott insulator is prepared and subsequently allowed to expand in the combined superlattice potential with $V_{752} = 42.0(3) E_{R,752}$ and $V_{532} = 3.7(1) E_{R,532}$ for 1 second. The constructive interference in the triangular lattice results in a deeper potential compared to the destructive interference present in the honeycomb lattice, resulting in a much denser cloud following the same period of expansion. The combination of these in-situ measurements uniquely determines the superlattice phase.

In principle, each lattice may be independently calibrated to give a full reconstruction of the potential in the plane of the atoms. However, owing to limited power available in the 1D lattice, independent calibration with modulation spectroscopy is difficult as the band transitions are not truly resolved. Instead, precise knowledge of the depth of the square lattice, $V_{752}$, combined with knowledge of the relative tunnelings (obtained from correlation maps of the system), can be used to obtain the 1D lattice depth. We empirically find the depth of the 1D lattice that equalises $t_x, t_y$ with the diagonal tunneling $t_d$. This is done by experimentally equalising the nearest-neighbor two-point spin correlations in the triangular lattice. The depth of the square lattice used in the experiment is measured to be $V_{752} = 2.9(1)E_{R,752}$. At the point where we obtain an isotropic triangular lattice connectivity, we infer the depth of the 1D lattice using the computed band structure to be $V_{532} = 6.7(2)E_{R,532}$. This corresponds to absolute tunneling strengths of $t_x = t_y = t_d  =h \times  400(20)$ Hz. 

\subsection*{Full spin-charge readout in a bilayer imaging scheme}

\begin{figure}[]
\includegraphics[width = \columnwidth]{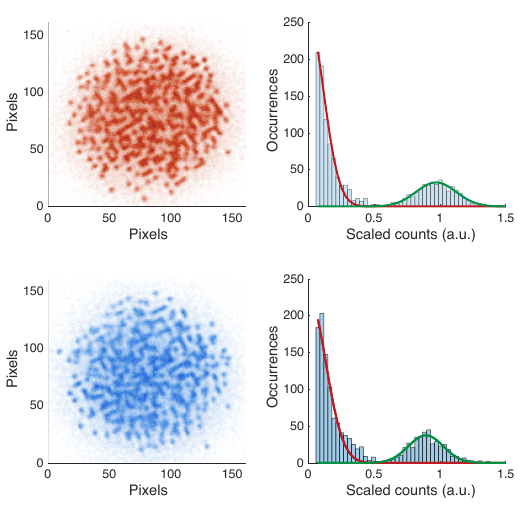}
\caption{\label{fig:Reconstruction} \textbf{Bilayer Image Reconstruction.} Sample deconvolved experimental images and occupation histograms for state $\ket{3}$ (top, red) and state $\ket{2}$ (bottom, blue) atoms. We use the Lucy-Richardson algorithm with five iterations for the deconvolution.
}
\end{figure}

Simultaneous imaging of charge and spin information is performed using a bilayer imaging scheme using Raman sideband cooling, similar to the method discussed in a prior publication~\cite{Yan2022}.  Minor differences from the prior scheme are discussed here.

Imaging consists of four steps:  
\begin{enumerate}
\item Tunneling is quenched by deepening the 2D lattice depth to $56.3(4)E_{R,752}$ in 170$\mu$s.  The axial confinement lattice is turned off in 20\,ms.  Atoms in the ground hyperfine state $|1\rangle$ are transferred to hyperfine state $|2\rangle$ using a radiofrequency Landau-Zener sweep lasting 50\,ms. In this state, the magnetic moment is of opposite sign to state $|3\rangle$. The magnetic Feshbach field is turned off in 10\,ms. 
 \item A magnetic field gradient of 336\,G/cm is applied to separate the two spin components $|2\rangle$ and $|3\rangle$ along the $z$-axis.  For this step, the 2D lattice is increased to $160(1)E_{R,752}$. 
\item Each spin component is trapped by a light sheet potential with vertical waist $w_z \approx 5$ $\mu$m which is turned on in 20~ms. The two potentials are moved further apart in the $z$-direction using a minimum-jerk trajectory in 4\,ms to a final separation of 16\,$\mu$m.  
\item Both layers are imaged simultaneously using Raman sideband cooling over a 2\,s duration.  The fluorescence photons are collected by a microscope objective and focused on two separate areas of a CMOS camera.
\end{enumerate}

\begin{figure*}[]
\includegraphics[width = 0.8\textwidth]{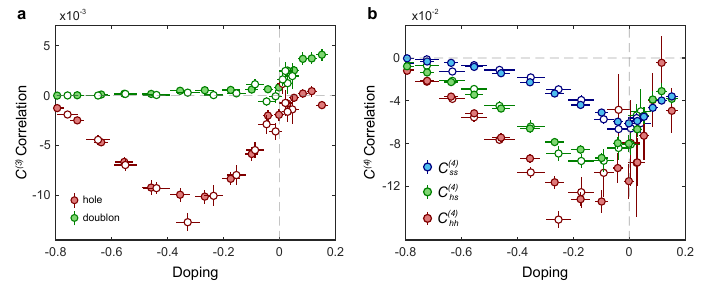}
\caption{\label{fig:varyChemPotential} \textbf{Multi-point Correlations at Different Global Chemical Potentials.} \textbf{a,} $C^{(3)}_h$ (red data) and $C^{(3)}_d$ (green data) evaluated at the bond closest to the dopant, and \textbf{b,} $C^{(4)}$ for two different datasets with distinct global chemical potentials.  As these two datasets track each other well, particularly at low filling, we conclude spatial gradients do not appreciably suppress resonant tunneling to affect measured correlations. Filled data points have mean atom number 799(35) while empty data points have mean atom number 622(27).}
\end{figure*}

\subsection*{Imaging fidelity}
During the imaging procedure described above, various types of errors can accumulate that will affect imaging fidelities.  RF spin-flip fidelities exceed 99\%, and these errors play a negligible role compared to other infidelities.  

Transport fidelity encompasses various errors that occur during the vertical motion of the atoms. First, we observe some atom loss that could be due to off-resonant scattering or background gas collisions. A surviving atom may hop to other sites of the same layer, which disturbs magnetic correlations or, if the final site is already occupied, leads to atom loss due to parity imaging. Finally, atoms may be transported into the wrong layer, so that they are assigned to the wrong spin state.

These effects are difficult to isolate and characterize independently.  We instead benchmark a related quantity: by preparing an almost unity-filled Mott insulator, we observe the proportion of singly occupied sites with and without the transport step. We prepare Mott insulating states with 97.1(4)\% singles fraction, verified by imaging both spin states in a single layer.
Any sites with zero atoms or two atoms appear dark from parity imaging. The transport step is tested by adding the Stern-Gerlach and optical transport (steps 2-3), and then reversing those steps to transport both spin states back into a single layer.  Then, the visible singles fraction drops to 95.4(5)\%. We assume that transport hopping errors populate randomly distributed sites and are irreversible. A hopping event will create a hole and a double occupancy. Therefore, this test indicates a transport infidelity of at most 0.9(3)\%. 

Additionally, errors may accrue during the Raman sideband cooling, appearing as loss (3.9\%), interlayer hopping (0.5\%), and intralayer hopping (negligible).  

Finally, errors can be introduced during image processing when we digitize the images into an occupancy matrix.  Compared to the bilayer readout of a sparse tweezer array of fewer than 50 atoms~\cite{Yan2022}, our current bilayer imaging scheme must reliably reconstruct atomic distributions of hundreds of atoms with high filling. Each layer adds an out-of-focus background on the image of the opposite layer, decreasing our signal-to-noise.  We choose a 2\,s Raman imaging time as a compromise between increasing the ratio between the desired signal and the background layer noise and minimizing hopping and loss errors. The problem of the out-of-focus background ultimately limits the peak densities we can reliably probe.  Empirically, we find that beyond dopings of 0.2, distinguishing between empty and occupied sites becomes difficult. We include a representative image of the occupation histograms for both imaging layers and the corresponding Gaussian fits to the zero and single atom peaks in Extended Data Fig. \ref{fig:Reconstruction}.


\subsection*{Calculation of Correlation Functions}
The experimental correlation functions presented in the text are computed as the fully connected three-point correlation function of a three-observable operator:

\begin{align}
    &\langle {\hat{n}^h_i \hat{S}^z_j \hat{S}^z_k} \rangle_c \equiv \langle (\hat{n}^h_i - \langle \hat{n}^h_i \rangle ) ( \hat{S}^z_j - \langle \hat{S}^z_j \rangle )  ( \hat{S}^z_k - \langle \hat{S}^z_k \rangle ) \rangle \nonumber \\
   &\quad   =\langle {\hat{n}^h_i \hat{S}^z_j \hat{S}^z_k} \rangle - \langle \hat{n}^h_i \rangle \langle \hat{S}^z_j \hat{S}^z_k \rangle - \langle \hat{n}^h_i  \hat{S}^z_j \rangle \langle \hat{S}^z_k \rangle \nonumber \\& \quad \quad  -\langle \hat{n}^h_i  \hat{S}^z_k \rangle \langle \hat{S}^z_j \rangle  + 2 \langle \hat{n}^h_i \rangle \langle \hat{S}^z_j \rangle \langle \hat{S}^z_k \rangle 
\end{align}
In particular, we do not \emph{a priori} assume a perfectly-spin balanced gas, compared to the simplified definition in Eq. 3 of the main text. 

As we equalise the tunnelings in all three directions and the same lattice depths are used for all datasets, we average over all 120$\degree$ and reflection symmetric higher-order correlators for plots versus doping. In addition, all DQMC calculations are done with the assumption that $t_x = t_y = t_d$. For completeness, in Extended Data Fig. \ref{fig:NonSymmetric_Snowflake} and Extended Data Fig. \ref{fig:NonSymmetric}, the same data as Fig. 2a. of the main text is shown without symmetrisation. There appear to be no major systematic differences between correlators of different orientations. This feature holds for all datasets and three- and four-point correlators shown in the main text.

\begin{figure}[t]
\begin{center}
\includegraphics[width=\columnwidth]{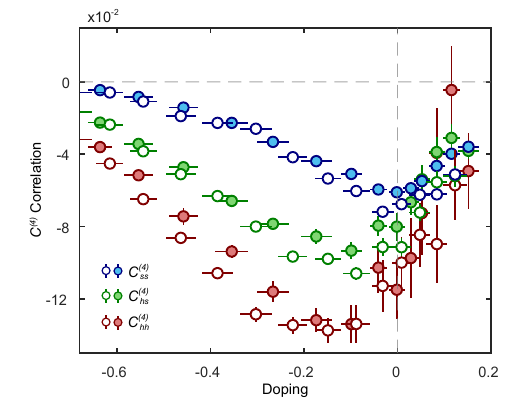}
\caption{\label{fig:fourpointratio} \textbf{Interaction Dependence of Four-point Correlations.} Four-point correlations vs. doping for different interaction strengths $U/t = 8.0(2)$ (open circles) and $U/t = 11.8(4)$ (filled circles).}
\end{center}
\end{figure}

\subsection*{Bootstrapping Error Analysis}
We use a bootstrapping analysis technique to obtain vertical errorbars for the $U/t = 11.8$ (1,146 experimental runs), $U/t = 8.0$ (535 experimental runs) and $U/t = 4.4$ (360 experimental runs) datasets for all correlators. The experimental runs are randomly separated into 80 groups, and the relevant three- and four-point correlations are calculated for each group. We sample from these 80 groups with replacement 10,000 times to obtain 10,000 bootsamples. We average over all lattice symmetries before taking the standard deviation of the bootsamples.

\subsection*{Effects of Spatial Gradients}

To create strongly interacting samples with a high central density, additional radial confinement is provided by an external dipole trapping beam at 1070 nm propagating approximately orthogonal to the atom plane. For the most strongly interacting datasets at $U/t \sim 12$, the radial trap frequency $\omega_r$ is approximately $2 \pi \times 370$ Hz. While the gradient near the center of the trap (regions of highest density) remains small, gradients away from the center of the lattice have the potential to impact resonant tunneling and affect correlations, particularly long-range and multi-point correlations. We empirically test for such by comparing two datasets with different global chemical potentials (different total atom number) but otherwise identical science parameters \hbox{\cite{koepsell2020microscopic}}. A global chemical potential shift will displace a bin of given density to a different radial position and hence cause it to sample a different spatial gradient. Disagreement between the two sets, particularly at low densities where the gradient is largest, would therefore indicate an effect due to the spatial gradient. In Extended Data Fig. {\ref{fig:varyChemPotential}}, we show measured three- and four-point correlations vs. doping for two datasets with different total atom number. We find no significant systematic deviations between these two datasets, from which we conclude that the gradient at the level present in the experiment does not affect measured correlation functions within experimental error bars.

\subsection*{Four-Point Correlations Interaction Dependence}
Four-point correlations displayed in Fig. 4 of the main text are displayed for the strongest interacting sample. In Extended Data Fig.~{\ref{fig:fourpointratio}} we compare these measurements to the more weakly interacting dataset at $U/t = 8$ to probe the evolution with interaction strength. Close to half filling we do not measure a significant difference in the strongest correlation $C^{(4)}_{hh}$, while the two correlations with singlon nearest neighbors $C^{(4)}_{hs}$ and $C^{(4)}_{ss}$ are reduced in magnitude with increasing interactions. This is consistent with kinetic magnetism being enhanced relative to super-exchange. Nonetheless, the variable temperature between datasets makes quantitative comparison difficult.

\begin{figure}[]
\includegraphics[width = \columnwidth]{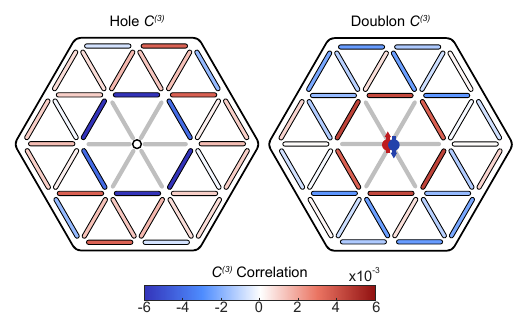}
\caption{\label{fig:NonSymmetric_Snowflake} \textbf{Unsymmetrised Correlations versus Distance.}  $C^{(3)}_h$ for $\delta=-0.10(2)$ and $C^{(3)}_d$ for $\delta =0.15(2)$ out to $d = (2,0)$ without averaging over the 120 degree rotational symmetry.}
\end{figure}

\begin{figure*}[]
\includegraphics[width = 0.6\textwidth]{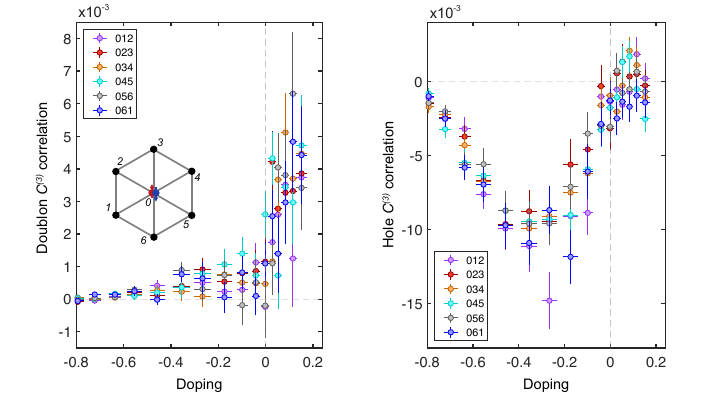}
\caption{\label{fig:NonSymmetric} \textbf{Unsymmetrised Correlations versus Doping.} $C^{(3)}_h$ and $C^{(3)}_d$ for different dopings at a $U/t = 11.8$ without averaging over all six individual plaquettes. We see that unsymmetrised correlations are largely consistent between different orientations. Errorbars are 1 s.e.m.}
\end{figure*}

\subsection*{Solution of the Hubbard Model on a Single Plaquette}

In this section we present the ground state wavefunctions and correlation functions on a three-site triangular plaquette. This toy example displays many of the important features that are present in the larger lattice system and is therefore instructive to consider in detail. The ground state of the triangular plaquette above zero doping with $N = 4$ spin has energy $E_{N = 4} = -2t + U$  in all magnetization sectors $S_z \in \{-1, 0, 1\} $ for $U > 0$, with ground state wavefunctions given by \cite{Merino_Powell_McKenzie_2006}:

\begin{align}
    \ket{\psi^1_{N = 4}} = &\frac{1}{\sqrt{3}} \big( \ket{\uparrow \downarrow,\downarrow,\downarrow} - \ket{\downarrow,\uparrow \downarrow,\downarrow} + \ket{\downarrow,\downarrow,\uparrow \downarrow} \big) \nonumber \\
    \ket{\psi^0_{N = 4}} = &\frac{1}{\sqrt{6}} \big[ \big( \ket{\uparrow \downarrow, \uparrow,\downarrow} + \ket{\uparrow \downarrow, \downarrow,\uparrow} \big) \nonumber\\
    - \big( &\ket{ \uparrow,\uparrow  \downarrow,\downarrow} + \ket{ \downarrow,\uparrow \downarrow,\uparrow} \big)\nonumber  + \big( \ket{ \uparrow,\downarrow,\uparrow \downarrow} + \ket{ \downarrow,\uparrow,\uparrow \downarrow} \big)  \big] \nonumber \\
    \ket{\psi^{-1}_{N = 4} } = &\frac{1}{\sqrt{3}} \big( \ket{\uparrow \downarrow,\uparrow,\uparrow} - \ket{\uparrow,\uparrow \downarrow,\uparrow} + \ket{\uparrow,\uparrow,\uparrow \downarrow} \big).
\end{align}
The spin wavefunction is in the symmetric (triplet) configuration with the doublon delocalized across the triangular plaquette. The basis independent spin-spin correlation function $\langle\hat{\mathbf{S}}_i \cdot \hat{\mathbf{S}}_j \rangle= 1/3$ indicates ferromagnetic alignment in all bases, with the fermionic spin operators defined as:
\begin{equation}
    \hat{\mathbf{S}}_i = \hat{c}^\dag_{i \alpha} \boldsymbol{\sigma}_{\alpha \beta} \hat{c}_{i \beta},
\end{equation}
where $\alpha, \beta \in \{ \uparrow, \downarrow\}$ and $\boldsymbol{\sigma} = ( \sigma^x, \sigma^y,\sigma^z)$ are the Pauli matrices.

Below half-filling, i.e. $N = 2$, magnetization sectors are no longer degenerate at any value of $U$ and the ground state has $S_z = 0$. In this case, the ground state is:

\begin{align}
    \ket{\psi^0_{N = 2} } = &\alpha \big[\big( \ket{\circ, \uparrow, \downarrow} - \ket{\circ, \downarrow,\uparrow} \big)   + \big(\ket{\uparrow, \circ, \downarrow} - \ket{\downarrow, \circ, \uparrow}\big)\nonumber\\
    &+ \big(\ket{\uparrow , \downarrow, \circ} - \ket{\downarrow, \uparrow, \circ}\big)\big] \nonumber \\
     &-  \beta  \big(\ket{\uparrow \downarrow, \circ, \circ} + \ket{\circ,\uparrow \downarrow, \circ} + \ket{\circ, \circ, \uparrow \downarrow}\big),
\end{align}

with energy 

\begin{align}
    E^0_{N = 2}  = \frac{-2 t + U - \sqrt{36 t^2 + 4 t U + U^2}}{2},
\end{align}
where $\alpha = \mu/\sqrt{6(1 + \mu^2)}$, $\beta = 1/\sqrt{3(1 + \mu^2)}$ and 

\begin{align}
    \mu = -\frac{\sqrt{36 t^2+4 t U+U^2}+2 t+U}{4 \sqrt{2} t}.
\end{align}
In particular, the limiting cases of infinite and vanishing interactions are straightforward: $\beta \to 0, \alpha \to 1/\sqrt{6}$ as $ U \to \infty$ and $\alpha, \beta \to 1/3$ as $U \to 0$. We may therefore compute relevant experimental correlations functions for this simple example in all cases.

The theoretical correlations shown in Table \ref{tab:one} assume equal weights over all degenerate ground states for $ \delta = +1/3$, as although global magnetisation in the lattice is conserved, local magnetization on a single plaquette may fluctuate as the bulk acts as a particle bath.

\begin{table}
\begin{center}
    \begin{tabular}{|c|c|c|}
         \hline 
          & $\delta = -1/3$ & $\delta = +1/3$\\
          \hline
           $U \to \infty$ &$\begin{aligned} \\ \langle{n^h}\rangle &= 1/3 \\ \expval{S^z S^z} &= -1/3 \\  \langle{n^hS^z S^z}\rangle &= -1/3 \\ \langle{n^h S^z S^z}\rangle_c &= - 2/9 \\ \phantom{text} 
           \end{aligned}$  & $\begin{aligned} \\ \langle{n^d}\rangle &= 1/3 \\ \langle{S^z S^z}\rangle &= 1/9 \\  \langle{n^dS^z S^z}\rangle &= 1/9 \\ \langle{n^d S^z S^z}\rangle_c &=  2/27 \\ \phantom{text} \end{aligned}$  \\ 
           \hline
           $U  = 0$ & $\begin{aligned} \\ \langle{n^h}\rangle &= 4/9 \\ \langle{S^z S^z}\rangle &= -2/9 \\  \langle{n^h S^z S^z}\rangle &= -2/9 \\ \langle{n^h S^z S^z}\rangle_c &= - 10 / 81 \\ \phantom{text} \end{aligned}$ &$\begin{aligned} \\  \langle{n^d}\rangle &= 10/27 \\ \langle{S^z S^z}\rangle &= 4/27 \\  \langle{n^d S^z S^z}\rangle &= 4/27 \\ \langle{n^d S^z S^z}\rangle_c &= 68/729 \\ \phantom{text} \end{aligned}$ \\
           \hline
           
    \end{tabular}
    \end{center}
    \caption{\label{tab:one}\textbf{Single plaquette correlation functions.}}
    \end{table}

\newpage
\subsection*{Theory comparison with DQMC}
We use the QUEST package ~\cite{varney2009quantum} to calculate the theoretical correlations in the triangular Fermi-Hubbard model implemented in this work.  QUEST is a Fortran-based package using Determinant Quantum Monte Carlo (DQMC) to study many-body problems with unbiased numerical methods.  We point out that DQMC can fail to converge at low temperatures and large interactions strengths, but in the parameter regimes probed by this study ($T/t > 0.5$, $U/t < 12$), the fermion sign problem does not prevent results from reliably converging.

The simulations are run on an 8x8 isotropic triangular lattice with the inverse temperature split
into $L = 80$ imaginary time slices, where the inverse temperature $\beta = L \delta\tau$. We perform 5000 warmup sweeps, 20000 measurement sweeps, and 200 bins for statistics using Princeton University's Della cluster.

\begin{figure}[]
\includegraphics[width = \columnwidth]{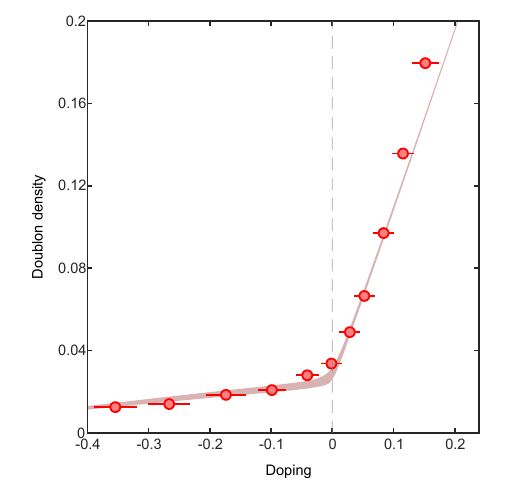}
\caption{\label{fig:Doubles} \textbf{Doublon Density versus Doping.} Number of measured doubles (red points) and theoretical expected number of doubles from DQMC (red band) with imaging fidelity of 0.96 accounted for at $U/t = 11.8(4)$, $T/t = 0.94(4)$. The highest doping bin ($\delta = 0.15$) has around 20 percent more, the second highest doping bin ($\delta = 0.12$ has around 10 percent more, and the third highest doping bin ($\delta = 0.08$) has around 5 percent more doubles than predicted. We believe this is caused by image reconstruction errors and leads to a systematic underestimate of certain three and four point correlators above zero doping. This qualitative trend appears for all three interaction strengths. Error bars are 1 s.e.m.}
\end{figure}

\begin{figure}[]
\includegraphics[width = \columnwidth]{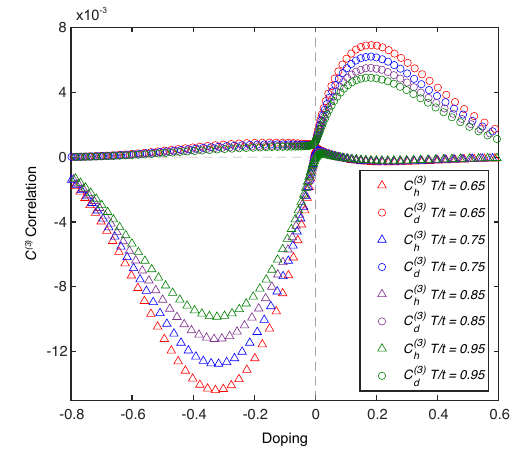}
\caption{\label{fig:Temperature} \textbf{Three-point Correlations Temperature Dependence.} DQMC results for $C^{(3)}$ evaluated at the closest bond to the dopant versus doping at a fixed $U/t = 12$ and different temperatures. All temperature curves have the same qualitative trend with the $C^{(3)}_h$ minimum at a doping of $\sim -0.3$ and the $C^{(3)}_d$ maximum at a doping of $\sim 0.15$ with a linear region near zero doping. Decreasing $T/t$ of the gas from $0.95$ to $0.65$ causes the magnitude of the peak values of $C^{(3)}$ to increase by roughly fifty percent. An imaging fidelity of 0.96 is assumed.
}
\end{figure}

\begin{figure*}[]
\begin{center}
\includegraphics[width=0.7\textwidth]{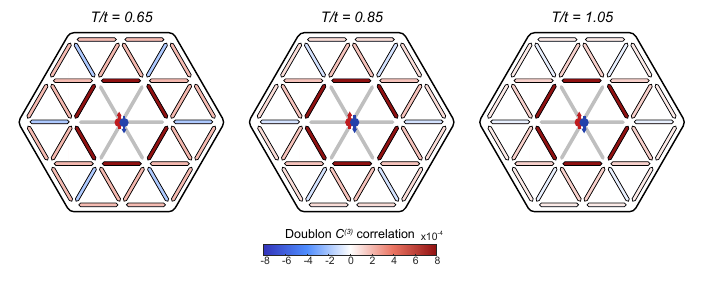}
\caption{\label{fig:structureVsTemp} \textbf{Doublon Correlations versus Temperature.} $C^{(3)}_d$ calculated at $\delta = 0.02$ for $U/t$ = 12 at three different temperatures using DQMC. As the temperature decreases, the magnitude of the farther correlations increases slightly while there are no major qualitative differences in the structure of the correlations; the size of the polaron does not have a strong dependence on the temperature.}
\end{center}
\end{figure*}

\subsection*{Three- and four- point correlators in DQMC}
For completeness, we detail how the three- and four-point correlators are numerically obtained in DQMC.  The single-particle Green's function determines the DQMC dynamics and is used to compute observables.  Using Wick's theorem, arbitrary correlations can be computed~\cite{assaad2022}.  

For example, a bare three-point correlator $\langle \hat{n}^s_i~ \hat{n}^t_j~ \hat{n}_k^u\rangle$ of densities on three distinct sites ($i,j,k$), hosting spin labels ($s,t,u \in \{ \uparrow, \downarrow \})$ respectively, can be written as the determinant
\[
\begin{vmatrix}
1-G_s(i,i) & -G_t(i,j)\delta_{s,t} & -G_u(i,k)\delta_{s,u}\\
-G_s (j,i)\delta_{s,t} & 1-G_t (j,j) & -G_u (j,k) \delta_{t,u} \\
-G_s(k,i)\delta_{s,u} & -G_t(k,j)\delta_{t,u} & 1-G_u(k,k)
\end{vmatrix}
\]
where $G_\sigma(i,j) = \langle \hat{c}_i^\sigma \hat{c}^{\dagger \sigma}_{j}\rangle$ is the single particle Green's function.  Extensions to four-, five-, and six- point density correlators can easily be written.
Thus, data involving the hole-spin-spin correlators were compared to DQMC numerical results by first defining this operator as 
\begin{equation}
    C^{(3)}_{h,bare}(i,j,k) = \nonumber \expval{(1-\hat{n}_i^\uparrow-\hat{n}_i^\downarrow+\hat{n}_i^\uparrow\hat{n}_i^\downarrow)(\hat{n}_j^\uparrow-\hat{n}_j^\downarrow)(\hat{n}_k^\uparrow-\hat{n}_k^\downarrow)}
\end{equation}
and using a symbolic algebra
system to enumerate the Wick contractions and automatically generate measurement code for this observable. The correlator $C^{(3)}_h$ involves up to eight fermion operators, whereas correlators such as $C^{(4)}_{hh}$ involve up to twelve.

\subsection*{Fitting temperatures and interaction strengths}

To determine $U/t$ and $T/t$ for each dataset, we do a least-squares fit using the measured densities ($n^h$, $n^s$, $n^d$) and the nearest-neighbor single-spin two-point correlation $C_{\uparrow} = \langle\hat{n}_i^\uparrow \hat{n}_j^\uparrow  \rangle - \langle\hat{n}_i^\uparrow  \rangle \langle\hat{n}_j^\uparrow  \rangle$  as a function of doping to interpolated DQMC functions. Lattice sites are grouped by doping such that each datapoint represents approximately the same number of lattice sites. Groupings are approximately radial but reflect the slight ellipticity and asymmetry of the atom distribution in the lattice. Error bars on the doping come from the standard deviation of the average doping for each lattice site within a given grouping. Errorbars for $U/t$ and $T/t$ come directly from the least-squares fits. Additionally, we make the overall imaging fidelity a free parameter in the least-squares fit. For the $U/t= 4.4, 8.0,$ and $11.8$ datasets we fit imaging fidelities of 0.957(4), 0.954(3), and 0.951(3) respectively, consistent with the typical bilayer fidelity. The experimental doping in all plots is the measured value scaled by the loss imaging fidelity of 0.96. All theory curves for three-point and four-point correlators are DQMC results corrected for the loss imaging fidelity. For example, for an imaging fidelity $1-\epsilon$, the appropriate connected three-point correlator, corrected for imaging fidelity, is
\begin{equation}
    C^{(3)}_{h} = (1-\epsilon)^2 C^{(3)}_{h,\mathrm{DQMC}} + (1-\epsilon)^2 \epsilon C^{(3)}_{s,\mathrm{DQMC}}, 
\end{equation}
where $C^{(3)}_{h,\mathrm{DQMC}}$ is the uncorrected DQMC output for the connected three-point hole-spin-spin correlator, $C^{(3)}_{s,\mathrm{DQMC}}$ is the connected three-point singlon-spin-spin correlator. The $(1-\epsilon)^2$ in the first term corrects for the potential loss of the spins during imaging, while the second term corrects the possibility that the measured hole was due to a single atom that was lost during imaging. Higher order corrections are much smaller at these interactions, temperatures and dopings for $\epsilon = 0.04$. For $C^{(3)}_{d}$, the correction is
\begin{equation}
    C^{(3)}_{d} = (1-\epsilon)^4 C^{(3)}_{d,\mathrm{DQMC}}.
\end{equation}
The DQMC theory fits the data well except for densities above half filling, where we systematically overstimate the number of doubles and holes and underestimate the number of singles for a given doping in all datasets (Extended Data Fig.~\ref{fig:Doubles}). We attribute this deviation to the onset of an increase in reconstruction errors with increasing particle doping, where we likely identify false positive atoms in a given imaging layer due to the background signal from the other imaging layer. This systematic error is not taken into account directly in any imaging fidelity or errorbar in the main text.

\subsection*{Theoretical three-point correlators vs. interaction strength and temperature}

\begin{figure}[]
\includegraphics[width = \columnwidth]{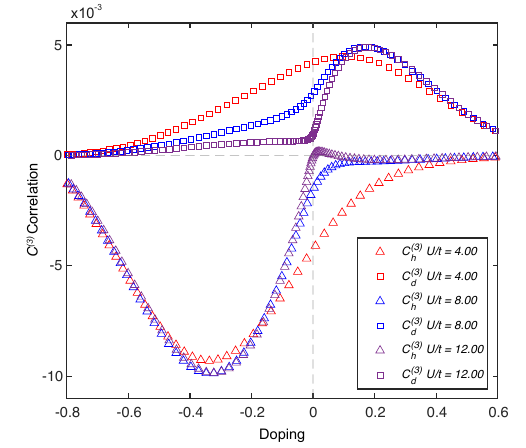}
\caption{\label{fig:Interaction} \textbf{Three-point Correlations Interaction Dependence.} DQMC results for $C^{(3)}$ evaluated at the closed bond to the dopant versus doping at fixed $T/t = 0.85$ and different interaction strengths. We see that for all interaction strengths the $C^{(3)}_h$ minimum is at a doping of $\sim -0.3$ and the $C^{(3)}_d$ maximum is at a doping of $\sim 0.15$ for the two higher interaction strengths, while for $U/t = 4$ the peak appears slightly closer to zero doping. The two higher doping curves have roughly the same peak $C^{(3)}$ magnitudes, while for $U/t = 4$ the peaks are roughly 10 percent lower. However, we see that qualitatively the curves are quite different close to half filling, where as $U/t$ increases the onset of $C^{(3)}_h$ and $C^{(3)}_d$ with doping becomes sharper, leading to a region where the $C^{(3)}_h$ is linear below zero doping and $C^{(3)}_d$ is linear above zero doping. An imaging fidelity of 0.96 is assumed. \\ \\ 
}
\end{figure}

In Fig. 3 of the main text, we plot $C^{(3)}$ evaluated at the closest bond to the dopant versus doping for three different $U/t$. Although all three datasets are at slightly different temperatures, this does not change the qualitative trends that exist with increasing interaction strength. To emphasise this, in Extended Data Fig. \ref{fig:Temperature}, we plot the $C^{(3)}$ on the triangular plaquette for multiple temperatures at a fixed interaction, and qualitatively all curves are similar with the main quantitative difference being that lower temperatures lead to larger magnitude $C^{(3)}$ peaks. This is also confirmed in Extended Data Fig. {\ref{fig:structureVsTemp}}, where spatial correlations plotted as a function of temperature close to half filling do not show variations in the structure, rather a general decrease in correlator amplitude. To illustrate that the key qualitative trends depend on $U/t$, we furthermore plot the $C^{(3)}$ evaluated at the closest bond to the dopant for multiple interactions at fixed temperature in Extended Data Fig. \ref{fig:Interaction}. Here, we see all of the same trends displayed in Fig. 3 of the main text.

\section*{Data availability} Source data are provided with this paper and can be found in the Harvard Dataverse \cite{DVN_ATI1FG_2023}. All other supporting data are available from the corresponding author upon request.

\section*{Code availability} The code used in this manuscript is available from the corresponding author upon request.

\end{document}